# A Hybrid Approach for Optimizing Planar Triangular Meshes

Gang Mei, John C.Tipper
Institut für Geowissenschaften – Geologie
Albert-Ludwigs-Universität Freiburg
Freiburg im Breisgau, Germany
{gang.mei, john.tipper}@geologie.uni-freiburg.de

Nengxiong Xu
School of Engineering and Technology
China University of Geosciences (Beijing)
Beijing, China
xunengxiong@yahoo.com.cn

*Abstract*— Modified Direct Method (MDM) is an iterative scheme based on Jacobi iterations for smoothing planar meshes [4]. The basic idea behind MDM is to make any triangular element be as close to an equilateral triangle as possible. Based on the MDM, a length-weighted MDM is proposed and then combined with edge swapping. In length-weighted MDM, weights of each neighboring node of a smoothed node are determined by the length of its opposite edge. Also, the MDM, Laplacian smoothing and length-weighted MDM are all combined with edge swapping, and then implemented and compared on both structured and unstructured triangular meshes. Examples show that length-weighted MDM is better than the MDM and Laplacian smoothing for structured mesh but worse for unstructured mesh. The hybrid approach of combining length-weighted MDM and edge swapping is much better and can obtain more even optimized meshes than other two hybrid approaches.

## I. INTRODUCTION

In finite element analysis it is important to always use high quality meshes: low quality meshes lead to unreliable results. After creating meshes, they usually need to be improved using various methods. There are two main classes of such methods. One is called *clear-up* techniques, which alter the connectivity between elements to improve mesh quality. The other is called *mesh smoothing*, which does not alter the element connectivity but relocates the nodes to improve the mesh quality. The most popular smoothing method is Laplacian smoothing (LS) [1, 2], which repositions each node at the centroid of its neighboring nodes in each iteration step.

The Direct Method (DM) which is not an iterative method was proposed by B.Balendran [3]. In the method, a system of optimization equations are generated based on elements for all new nodal locations and then be solved for the constraints once to get the final smoothed coordinates for all nodes.

Modified Direct Method (MDM) is proposed by G.Mei [4], which is an iterative method based on Jacobi iterations. It accepts the basic idea of DM, which is to make any triangular element be as close to an equilateral triangle as possible; meanwhile, to make any quadrilateral element be as close to a square as possible. To do this, firstly element stiffness matrices are created based on the type of elements. The modified forms of element stiffness matrices are simpler than those of original DM. And then by assembling all element stiffness matrices, a system of Jacobi iteration equations can be formed, which is different from the optimization equations in the original DM. Finally, the smoothed nodal coordinates can be generated by repeating Jacobi iterations until no nodes are moved beyond a specified tolerance in a same step. It is an iterative scheme that all nodes in a mesh are relocated one by one.

In original MDM, when to make a triangular element be an equilateral triangle, each node of the triangle can have a moved location in the operation. For a node A which is shared by several, for example *n*, triangular elements, this node can have *n* optimized locations noted as $A_i^{new}$; its final position $A^{final}$ is the average location of those *n* moved locations:

$$A^{final} = \frac{1}{n}\sum_{i=1}^{n} A_i^{new}$$

The above equation can be represented as a weighed form: the weight factor $w_i$ of each moved location for node A is $1/n$,

$$A^{final} = \sum_{i=1}^{n}(w_i \cdot A_i^{new}), \text{ with } w_i = \frac{1}{n}$$

In length-weighted MDM, when adjust a triangular element to be an equilateral triangle, each node of the triangle has its optimal location which is affected by its opposite edge. For instance, for a triangle △ABC, the new locations for the three nodes are affected by edges BC, CA and AB, respectively. For a node A shared by *n* triangular elements, it has *n* optimal location and each location denoted as $A_i^{new}$ is affected by its opposite edge. The weight factor $w_i$ of each smoothed location for node A is no longer $1/n$ but a simple function in which the variables is the length of the relevant opposite edge $l_i$:

$$A^{final} = \sum_{i=1}^{n}(w_i \cdot A_i^{new}), \text{ with } w_i = l_i / \sum_{i=1}^{n} l_i$$

This paper is organized as follows. In Sect.2, the original MDM applied on planar triangular mesh is described in details. Sect.3 shows how length-weighted MDM is developed and then combined with edge swapping. The application of length-weighted MDM and the comparison with the original MDM, Laplacian smoothing are generated in Sect. 4; also, the hybrid approaches by individually combining the above three methods with edge swapping are implemented and compared. Finally, in Sect.5 the conclusions are given.

This research was supported by the Natural Science Foundation of China (Grant Numbers 40602037 and 40872183) and Fundamental Research Funds for the Central Universities of China.

## II. MODIFIED DIRECT METHOD

DM is a directly smoothing technique in the sense that the new locations of all the nodes in a mesh are evaluated at the same time, see [3]. MDM accepts the basic idea behind DM of trying to adjust each element into its best shape by making any triangular element be as close to an equilateral triangle as possible. Comparing with DM, MDM also needs to create element stiffness matrix and then assemble a Jacobi iteration matrix, rather than a system of optimization equations. The smoothed nodal coordinates are obtained by solving the Jacobi iterations equations recursively.

### A. Element stiffness matrix

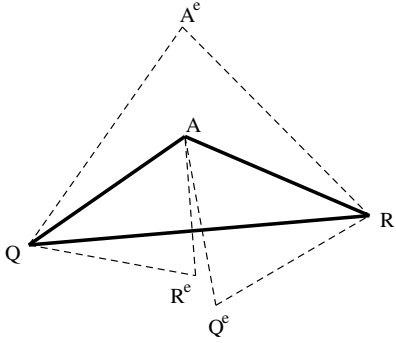

Figure 1. Optimal shape of a triangular element [3]

The optimal shape of a triangular element is equilateral triangle. Considering a triangular element with nodes A, Q and R as shown in Fig.1, if the nodes Q and R are fixed, let $A^e$ denote the new location of A, while Q, R and $A^e$ are the three vertices of an equilateral triangle. Hence, the coordinates of $A^e$ can be computed by Q and R as follows:

$$\begin{cases} X_A^e = \frac{1}{2}(X_Q + X_R) + \frac{\sqrt{3}}{2}(Y_Q - Y_R) \\ Y_A^e = \frac{\sqrt{3}}{2}(-X_Q + X_R) + \frac{1}{2}(Y_Q + Y_R) \end{cases} \quad (1)$$

, where $X_i$ and $Y_i$ denote the coordinates of node $i$. The equation (1) can be rewritten as (2).

$$\begin{cases} X_A^e = (0X_A + 0Y_A) + (\frac{1}{2}X_Q + \frac{\sqrt{3}}{2}Y_Q) + (\frac{1}{2}X_R - \frac{\sqrt{3}}{2}Y_R) \\ Y_A^e = (0X_A + 0Y_A) + (-\frac{\sqrt{3}}{2}X_Q + \frac{1}{2}Y_Q) + (\frac{\sqrt{3}}{2}X_R + \frac{1}{2}Y_R) \end{cases} \quad (2)$$

Similarly, $Q^e$ and $R^e$ that denote the new locations while A and R, A and Q are fixed, can also be calculated, see (3).

$$\begin{bmatrix} 0 & 0 & 1/2 & \sqrt{3}/2 & 1/2 & -\sqrt{3}/2 \\ 0 & 0 & -\sqrt{3}/2 & 1/2 & \sqrt{3}/2 & 1/2 \\ 1/2 & -\sqrt{3}/2 & 0 & 0 & 1/2 & \sqrt{3}/2 \\ \sqrt{3}/2 & 1/2 & 0 & 0 & -\sqrt{3}/2 & 1/2 \\ 1/2 & \sqrt{3}/2 & 1/2 & -\sqrt{3}/2 & 0 & 0 \\ -\sqrt{3}/2 & 1/2 & \sqrt{3}/2 & 1/2 & 0 & 0 \end{bmatrix} \begin{bmatrix} X_A \\ Y_A \\ X_Q \\ Y_Q \\ X_R \\ Y_R \end{bmatrix} = \begin{bmatrix} X_A^e \\ Y_A^e \\ X_Q^e \\ Y_Q^e \\ X_R^e \\ Y_R^e \end{bmatrix} \quad (3)$$

Equation (3) can be rewritten as Jacobi iteration equations, in which, $X_A$, $Y_A$, $X_Q$, $Y_Q$, $X_R$ and $Y_R$ are deemed as the values in $k$ step while $X_A^e$, $Y_A^e$, $X_Q^e$, $Y_Q^e$, $X_R^e$ and $Y_R^e$ can be considered as the values in $k+1$ step, see (4).

$$\begin{bmatrix} 0 & 0 & 1/2 & \sqrt{3}/2 & 1/2 & -\sqrt{3}/2 \\ 0 & 0 & -\sqrt{3}/2 & 1/2 & \sqrt{3}/2 & 1/2 \\ 1/2 & -\sqrt{3}/2 & 0 & 0 & 1/2 & \sqrt{3}/2 \\ \sqrt{3}/2 & 1/2 & 0 & 0 & -\sqrt{3}/2 & 1/2 \\ 1/2 & \sqrt{3}/2 & 1/2 & -\sqrt{3}/2 & 0 & 0 \\ -\sqrt{3}/2 & 1/2 & \sqrt{3}/2 & 1/2 & 0 & 0 \end{bmatrix} \begin{bmatrix} X_A^k \\ Y_A^k \\ X_Q^k \\ Y_Q^k \\ X_R^k \\ Y_R^k \end{bmatrix} = \begin{bmatrix} X_A^{k+1} \\ Y_A^{k+1} \\ X_Q^{k+1} \\ Y_Q^{k+1} \\ X_R^{k+1} \\ Y_R^{k+1} \end{bmatrix} \quad (4)$$

The left hand matrix in (4) is defined as the stiffness matrix of triangular element in 2D.

### B. Jacobi iteration matrix and equations

Considering a triangular mesh with $n$ nodes, each element has its stiffness matrix, and a $2n \times 2n$ Jacobi iteration matrix can be created by assembling all element matrices according to element connectivity. Meanwhile, a system of Jacobi iteration equations of the triangular mesh can be generated as (5), where $e_i$ represents the number of elements which share the node $i$.

$$\begin{bmatrix} \alpha_{11} & \alpha_{12} & \alpha_{13} & \cdots & \alpha_{1(2n)} \\ \alpha_{21} & \alpha_{22} & \alpha_{23} & \cdots & \alpha_{2(2n)} \\ \alpha_{31} & \alpha_{32} & \alpha_{33} & \cdots & \alpha_{3(2n)} \\ \vdots & \vdots & \vdots & \ddots & \vdots \\ \alpha_{(2n)1} & \alpha_{(2n)2} & \alpha_{(2n)3} & \cdots & \alpha_{(2n)(2n)} \end{bmatrix} \begin{bmatrix} X_1^k \\ Y_1^k \\ X_2^k \\ \vdots \\ Y_n^k \end{bmatrix} = \begin{bmatrix} e_1 \cdot X_1^{k+1} \\ e_1 \cdot Y_1^{k+1} \\ e_2 \cdot X_2^{k+1} \\ \vdots \\ e_n \cdot Y_n^{k+1} \end{bmatrix} \quad (5)$$

According to the iteration equations, the final coordinates of all nodes after smoothing can be generated by repeating the above system of equation in several iteration steps until a given tolerance is reached. The original coordinates of all nodes are considered as the values in the 0 step.

The MDM can be divided into three steps: firstly, search the elements which share a given node for each node; and then, assemble the element stiffness matrix of each element to form Jacobi iteration matrix; finally, repeat the iterations until a tolerance is reached to get final smoothed coordinates.

## III. THE HYBRID APPROACH

### A. Length-weighted Modified Direct Method

In original MDM, for a node such as A, its smoothed position at each iteration step is the average location of all $A^e$. This can be represented as following equation:

$$\sum (X_A - X_A^e) = 0, \quad \sum (Y_A - Y_A^e) = 0$$

, for example, in Fig.2, there is

$$\sum_{i=1}^{n} w_i (X_A - X_{Ai}^e) = 0, \quad \sum_{i=1}^{n} w_i (Y_A - Y_{Ai}^e) = 0$$

, where the weights $w_i = 1/n = 1/6$.

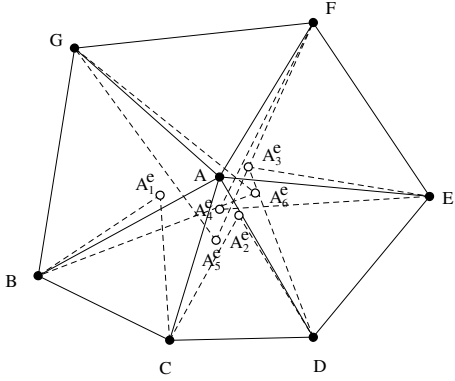

Figure 2. Nodal smoothed locations in triangular meshes

Considering effect of edges, $A_1^e$, $A_2^e$, $A_3^e$, $A_4^e$, $A_5^e$ and $A_1^e$ are relevant with the edges BC, CD, DE, EF, FG and GB, respectively (Fig.2). Let $l_i$ denote the length of edge $i$, then the weights $w_i$ can be computed by the edges, see (6).

$$w_i = l_i / \sum_i^n l_i, \quad \sum_i^n w_i = 1 \qquad (6)$$

For a triangular element, i.e., the one in Fig.1, the weight factors of three nodes $w_A$, $w_Q$ and $w_R$ can be determined by the edges QR, RA and AQ, respectively.

$$w_A = \frac{|QR|}{sum_A}, \quad w_Q = \frac{|RA|}{sum_Q}, \quad w_R = \frac{|AQ|}{sum_R} \qquad (7)$$

In (7), $sum_A$, $sum_Q$ and $sum_R$ represent the length of all related edges of node A, Q and R, respectively. For a triangular element, the length of all related edges of a node, such as $sum_A$ can be computed by adding the length of all related edges: $sum_A$ = |BC| + |CD| + |DE| + |EF| + |FG| + |GB|.

The length-weighted MDM can be also summarized into three main steps as that of MDM: firstly, search the elements which share a given node for each node; and then, assemble the Jacobi iteration matrix; and at last repeat the iterations until a tolerance is reached to obtain smoothed locations. Noticeably, the iteration matrix must be updated in each iteration step.

### B. Edge swapping

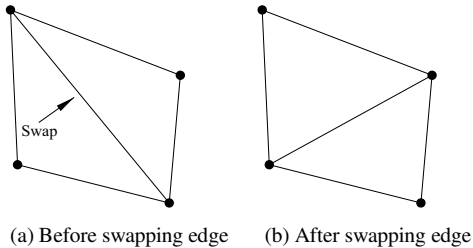

(a) Before swapping edge   (b) After swapping edge

Figure 3. Edge swapping

Edge swapping is one of clear-up operations for topological improvement [5]. It changes the shared edge of a pair of adjacent triangles without moving the nodes to improve the quality. Fig. 3(a) is an original pair of adjacent pairs and Fig. 3(b) is the new triangles after swapping the shared edge.

In this paper, minimum angle is accepted to decide whether a pair of adjacent triangles needs to swap edge. We compute the minimum angle of the original pair of triangles and the swapped pair of triangles. If the minimum angle of original pair of triangles is smaller than that of the new pair, edge swapping is necessary; otherwise, not.

### C. Hybrid approach

Firstly, triangular mesh is smoothed by length-weighted MDM, and in further edge swapping is accepted to alter the topology to improve the mesh quality.

## IV. EXAMPLES AND APPLICATIONS

In this section, we first give the criteria of measuring the quality of mesh, and then create several examples to test the performances of the hybrid algorithms.

### A. Quality of mesh

In order to test the performance of mesh improvement algorithms, the most basic method is to estimate the qualities of meshes before and after optimizing. A commonly used method is to check the quality of individual element and then the distribution of the qualities of all elements. Several methods for measuring element quality can be seen in [6] and [7].

For a triangular element, its optimal shape is equilateral triangle. Hence, its quality can be measured by how the triangle is close to an equilateral triangle. The triangular distortion metric proposed by Lee and Lo [8] is accepted in this paper. The distortion metric used for a triangular element with three nodes A, B and C can be computed according to (8).

$$\alpha = 2\sqrt{3}\frac{\|CA \times CB\|}{\|CA\|^2 + \|AB\|^2 + \|BC\|^2} \qquad (8)$$

The value $\alpha$ of an equilateral triangle is 1, and when the three nodes of a triangle element are collinear, the value $\alpha$ is 0. Hence, $\alpha$ value is between 0 and 1. The more a triangle element is close to an equilateral, the more the value $\alpha$ is close to 1.

### B. Structured triangular mesh

A simple structured triangular mesh is created as Fig. 4(a). Mesh optimized results are displayed in Fig. 4. Qualities of all elements in triangular mesh are tested according to (8). Also, average qualities is computed for each mesh, see Table 1.

Considering only smoothing, the smoothed meshed by Laplacian smoothing, MDM and length-weighted MDM are better than the original mesh. Noticeably, Laplacian smoothing and MDM has the same results, while length-weighted MDM is a little better than MDM.

Considering both mesh smoothing and edge swapping, the hybrid approaches by combining Laplacian smoothing with edge swapping, and combining MDM with edge have the same results. Since that edge swapping is implemented based on the same smoothed mesh. Comparing the above three combined approaches, it can be learnt from Table 1 that the combination of length-weighted MDM and edge swapping is much better than the rest of two.

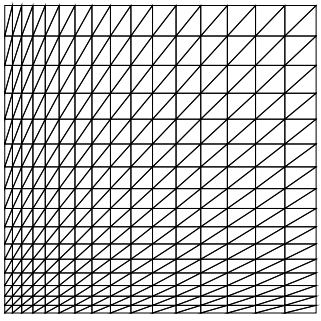

(a) Original

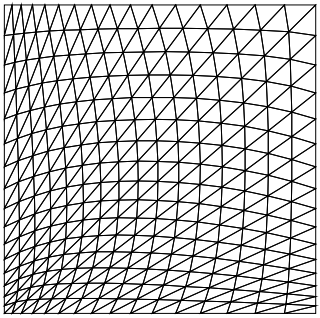   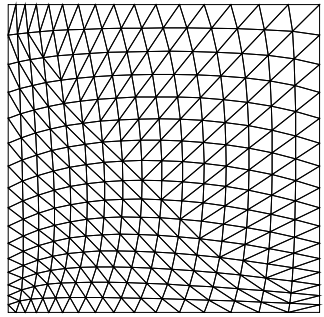

(b) By LS / MDM          (c) By (LS / MDM) + edge swapping

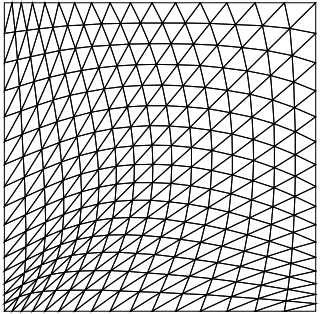   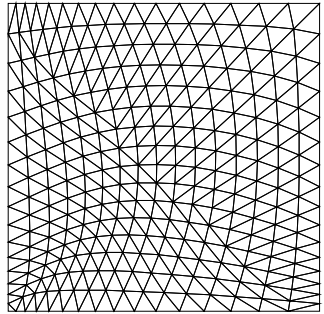

(d) By length-weighted MDM  (e) By length-weighted MDM + edge swapping

Figure 4.  Optimization for structured triangular mesh

TABLE I.        QUALITES OF OPTIMIZED MESHES

| Meshes | | Element quality (0.0~1.0) | | | | Average |
|---|---|---|---|---|---|---|
| | | *0.2~0.4* | *0.4~0.6* | *0.6~0.8* | *0.8~1.0* | |
| Fig.4 | (a) | 0.00% | 14.06% | 36.72% | 49.22% | 0.7555 |
| | (b) | 0.00% | 5.47% | 42.97% | 51.56% | 0.7910 |
| | (c) | 0.00% | 12.11% | 37.11% | 50.78% | 0.7957 |
| | (d) | 0.00% | 3.13% | 10.15% | 86.72% | 0.8866 |
| | (e) | 0.00% | 1.56% | 7.81% | 90.63% | 0.9101 |
| Fig.6 | (a) | 0.00% | 14.06% | 36.72% | 49.22% | 0.7555 |
| | (b) | 0.00% | 3.12% | 16.41% | 80.47% | 0.8845 |
| | (c) | 0.39% | 11.32% | 23.05% | 65.23% | 0.8355 |
| | (d) | 0.00% | 2.73% | 10.73% | 86.52% | 0.9026 |
| | (e) | 0.00% | 1.56% | 3.12% | 95.32% | 0.9216 |

From the equation (8), it can be concluded that the quality of a triangular element depends on its area and length of edges. Using the above two indicators, area and length of three edges, a scatter diagram can be drawn for a triangular mesh, in which, the horizontal axis represents area and vertical axis represents sum length of three edges of a triangle. Each point in the scatter diagram represents a triangular element.

If the point cloud in a scatter diagram is concentrated, it means that the triangular elements of a mesh nearly have the similar shapes and size; or in other words, they are even. The more the point cloud is concentrated, the more the triangular elements are even.

Fig. 5 are the scatter diagrams for the results by combining Laplacian smoothing with edge swapping, combining MDM with edge swapping and combining length-weighted MDM with edge swapping. The point cloud in Fig. 5(b) is much more concentrated than that in Fig. 5(a). This means the triangular mesh optimized by combined length-weighted MDM and edge swapping is better and much more even than those smoothed by other combined approaches.

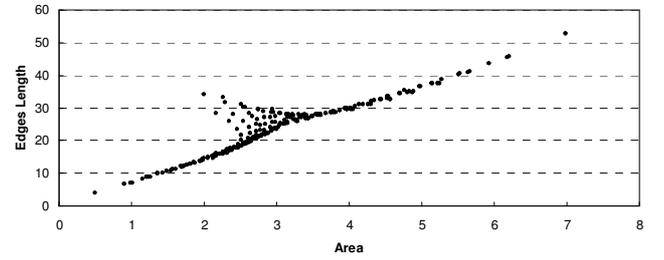

(a) By combining LS / MDM and edge swapping

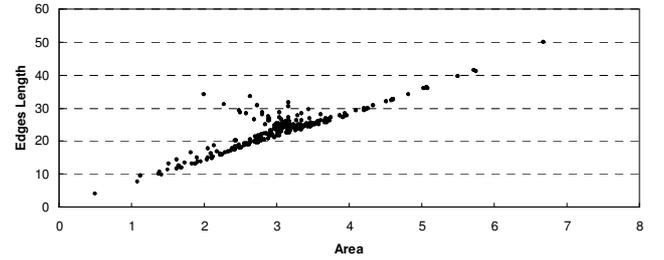

(b) By combining length-weighted MDM and edge swapping

Figure 5.  Scatter diagram of element quality of structured triangular mesh

### C. Unstructured triangular mesh

A simple triangular mesh has been created by Delaunay triangulation [9, 10], as shown in Fig. 6(a). Mesh optimized results are displayed in Fig. 6. Quality of each element of the unstructured triangular mesh is also evaluated according to the equation (8), and then all qualities are scattered.

When only the mesh smoothing is implemented, Laplacian smoothing and MDM also has the same results, while length-weighted MDM is worse than the MDM (Table1). Of the three combined approaches, the ones based on Laplacian smoothing and MDM also have the same results. And the method by combining length-weighted MDM with edge swapping is much better than the rest of two, as listed in Table 1.

Similar to that in Fig. 5, the point cloud in Fig. 7(b) is much more concentrated than that in Fig. 7(a). This means the mesh optimized by combining length-weighted MDM and edge swapping is much more even than those by other two.

An apparent limitation of the length-weighted MDM is that it is more expensive in time cost than the original MDM and Laplacian smoothing; since that it needs to compute and then assemble the iteration matrix in each iteration step.

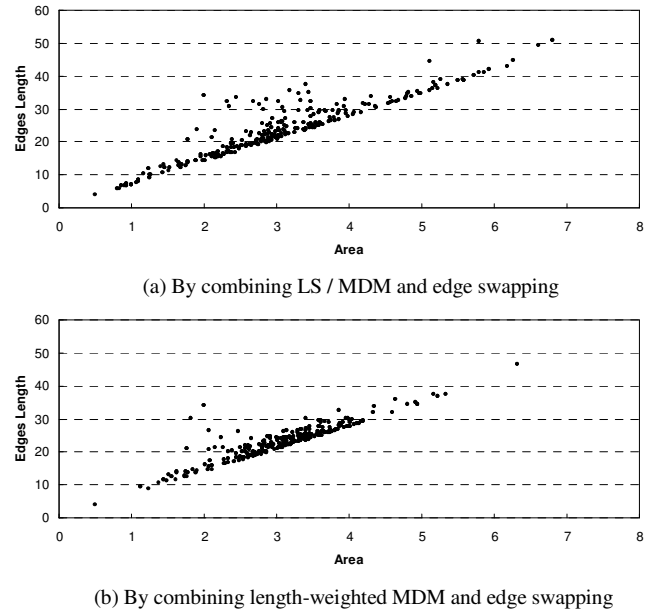

(a) By combining LS / MDM and edge swapping

(b) By combining length-weighted MDM and edge swapping

Figure 7. Scatter diagram of element quality of unstructured triangular mesh

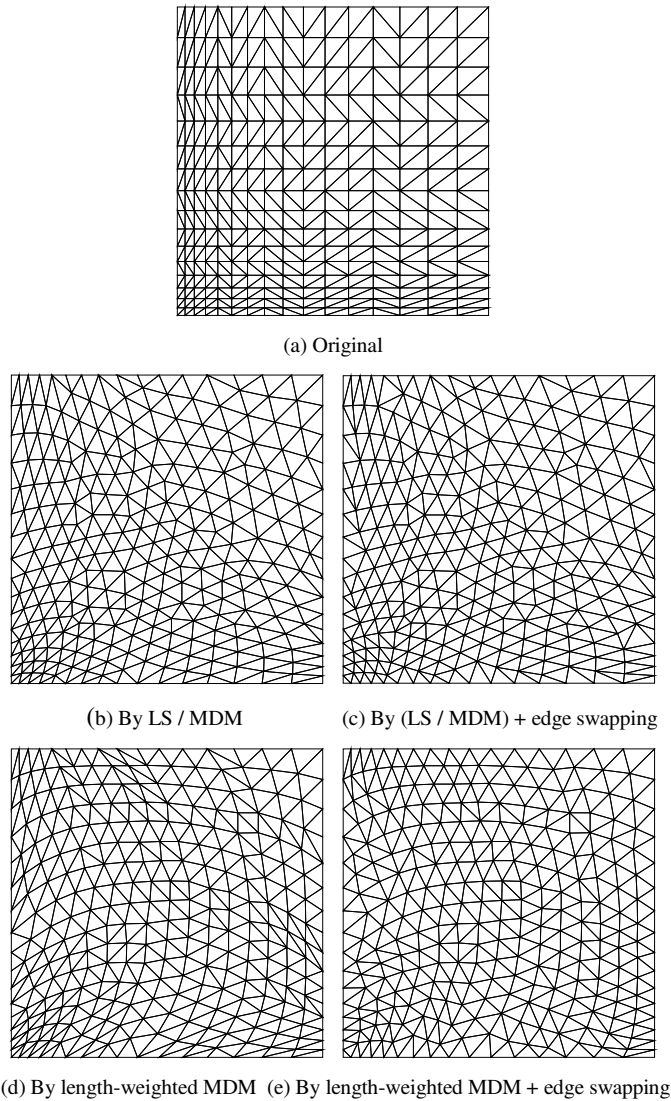

(a) Original

(b) By LS / MDM  (c) By (LS / MDM) + edge swapping

(d) By length-weighted MDM  (e) By length-weighted MDM + edge swapping

Figure 6. Optimization for unstructured triangular mesh

## V. CONCLUSION AND DISCUSSION

In this paper, we present a pure mesh smoothing algorithm, length-weighted MDM, and a hybrid approach by combining the length-weighted MDM with edge swapping. In the length-weighted MDM, weight factors for each neighboring node of a smoothed node are no longer the same but be computed by the length of related opposite edge.

Applying the length-weighted MDM on both structured and unstructured triangular meshes, it is better than the MDM and Laplacian smoothing for the structured mesh, but worse for the unstructured mesh. Applying combined length-weighted MDM and edge swapping on structured and unstructured meshes, it can obtain much better and more even elements than those by combining MDM or Laplacian smoothing with edge swapping.

ACKNOWLEDGMENT

The corresponding author Gang Mei would like to thank Chun Liu at Universität Kassel for sharing several ideas.

REFERENCES

[1] L. R. Herrmann, "Laplacian-isoparametric grid generation scheme," *Journal of the Engineering Mechanics Division,* vol. 102, pp. 749–907, 1976.

[2] S. A. Canann, J. R. Tristano, and M. L. Staten, "An approach to combined Laplacian and optimization-based smoothing for triangular, quadrilateral, and quad-dominant meshes," in *Proceedings of the 7th International Meshing Roundtable*, 1998.

[3] B. Balendran, "A direct smoothing method for surface meshes," in *Proceedings of the 8th International Meshing Roundtable*, 1999, pp. 189–193.

[4] G. Mei, "Modified Direct Method: an iterative algorithm for planar mesh smoothing," Submitted to *Computing and Visualization in Science,* 2012.

[5] W. H. Frey and D. A. Field, "Mesh relaxation: a new technique for improving triangulations," *International Journal for Numerical Methods in Engineering,* vol. 31, pp. 1121–1133, 1991.

[6] V. D. Liseikin, Grid Generation Methods, 2nd ed., vol. 1. Springer, 2010, pp.67–98.

[7] S. H. Lo, "A new mesh generation scheme for arbitrary planar domains," *International Journal for Numerical Methods in Engineering,* vol. 21, pp. 1403–1426, 1985.

[8] C. K. Lee and S. H. Lo, "A new scheme for the generation of a graded quadrilateral mesh," *Computers & Structures,* vol. 52, pp. 847–857, 1994.

[9] A. Bowyer, "Computing dirichlet tessellations," *The Computer Journal,* vol. 24, pp. 162–166, 1981.

[10] D. F. Watson, "Computing the n-dimensional Delaunay tessellation with application to Voronoi polytopes," *The Computer Journal,* vol. 24, pp. 167–172, 1981.